\documentclass[12pt]{article}

\usepackage{setspace}

\usepackage{amssymb}
\usepackage{amsfonts, amssymb,amsthm,amsmath, graphics,color,lscape,bm,float,graphicx, subfig,url}
\usepackage[colorlinks,citecolor=blue,urlcolor=blue]{hyperref}
\usepackage[numbers]{natbib}

\numberwithin{equation}{section}
\theoremstyle{plain}

\DeclareMathOperator*{\argmax}{arg\,max}

\newcommand{\bft} {\mbox{\boldmath $\theta$}}
\newcommand{\bfb} {\mbox{\boldmath $\beta$}}
\newcommand{\bfeta} {\mbox{\boldmath $\eta$}}
\newcommand{\bfe} {\mbox{\boldmath $\epsilon$}}

\newcommand{\e} {\mbox{\boldmath $e$}}
\newcommand{\Z}{{\mbox{\boldmath $Z$}}}
\newcommand{\Y}{{\mbox{\boldmath $Y$}}}
\newcommand{\y}{\mathbf{y}}

\title{\textbf{Modeling Nelson-Siegel Yield Curve using Bayesian Approach}}

\author{\textbf{Sourish Das\footnote{Infosys Foundation grants and TATA Trust grant to CMI partially supported Sourish Das's research. He is currently Commonwealth Rutherford Fellow at the S3RI of the University of Southampton.}}, \\
 Chennai Mathematical Institute, TN, INDIA \\
 S3RI, University of Southampton, UK
}

\begin{document}
\maketitle

\begin{abstract}
Yield curve modeling is an essential problem in finance. In this work, we explore the use of Bayesian statistical methods in conjunction with Nelson-Siegel model. We present the hierarchical Bayesian model for the parameters of the Nelson-Siegel yield function. We implement the MAP estimates via BFGS algorithm in \texttt{rstan}. The Bayesian analysis relies on the Monte Carlo simulation method. We perform the Hamiltonian Monte Carlo (HMC), using the \texttt{rstan} package. As a by-product of the HMC, we can simulate the Monte Carlo price of a Bond, and it helps us to identify if the bond is over-valued or under-valued. We demonstrate the process with an experiment and US Treasury's yield curve data. One of the interesting observation of the experiment is that there is a strong negative correlation between the price and long-term effect of yield. However, the relationship between the short-term interest rate effect and the value of the bond is weakly positive. This is because posterior analysis shows that the short-term effect and the long-term effect are negatively correlated. 
\end{abstract}


\section{Introduction} \label{Intro}

In financial applications, accurate yield curve modeling is of vital importance. Investors follow the bond market carefully, as it is an excellent predictor of future economic activity and levels of inflation, which affect prices of goods, stocks and real estate. The `yield curve' is a curve showing the interest rates across different maturity spans (1 month, one year, five years, etc.) for a similar debt contract. The curve illustrates the relationship between the interest rate's level (or cost of borrowing) and the time to maturity, known as the `term.' It determines the interest rate pattern, which you can use to discount the cash flows appropriately. The yield curve is a crucial representation of the state of the bond market. The short-term and long-term rates are usually different and short-term is lower than the long-term rates. The long-term rates are higher since the risk is more in long-term debt. The price of long-term bond fluctuates more with interest rate changes. The `term structure' tells us, at a given time, how the yield depends on maturity.  The most important factor in the analysis of the fixed-income asset is the yield curve. Any analysis of the fixed-income attribution requires evaluating how changes in the curve are estimated, and its impact on the performance of a portfolio. Some form of mathematical modeling of the yield curve is necessary, as it explains the curve's movement to extrapolate.

The slope of the yield curve is an essential indicator of short-term interest rates and is followed closely by investors \cite{yield_curve_imp}.  As a result, this has been the center of significant research effort. Several statistical methods and tools, commonly used in econometrics and finance, are implemented to model the yield curve (see for example, \cite{Nelson_Siegel_1987}, \cite{Diebold_Li_2006}, \cite{Hays_Shen_Huang_2012} and \cite{Chen_Niu_2014}).  The \cite{Nelson_Siegel_1987} introduces a parametrically parsimonious model for yield curves that has the ability to represent the shapes generally associated with yield curves; monotonic, humbed and $mathcal{S}$-shaped. 

Bayesian inference was applied to Dynamic Nelson-Siegel Model with Stochastic Volatility which models the conditional heteroscadasticity \cite{Joao.2010}. Bayesian inference for the stochastic volatility Nelson–Siegel (SVNS) model was introduced by \cite{Nikolaus.2012}. This models the stochastic volatility in the underlying yield factors. Bayesian extensions to Diebold-Li term structure model involve the use of a more flexible parametric form for the yield curve \cite{Laurini.2010}. It allows all the parameters to vary in time using a structure of latent factors, and the addition of a stochastic volatility structure to control the presence of conditional heteroskedasticity observed in the interest rates.

The Nelson-Selgel class of functions that produces the standard yield curve shapes associated with solutions to differential equations. If a differential equation produces the spot rates, then forward rates, being forecasts, will be the solution to the equations. Hence, the expectations theory of the term structure of the interest rates motivate investigating the Nelson-Siegel class.
For example, if the immediate forward rate at maturity $\tau$, denoted $r(\tau)$, is given by the solution to a second-order differential equation, with real and unequal roots, then we have 
$$
r(\tau)=\beta_0+\beta_1.\exp(-\tau/\lambda)+\beta_2.[(\tau/\lambda). \exp(-\tau/\lambda)].
$$
The yield to maturity on a bond, denoted by $\mu(\tau)$, is average of the forward rates
$$
\mu(\tau)=\frac{1}{\tau}\int_{0}^{\tau}r(\tau)d\tau,
$$
the resulting function is popularly known as the Nelson-Siegel function \cite{Nelson_Siegel_1987}, which has the form 
\begin{equation} \mu(\tau)=\beta_0+(\beta_1+\beta_2)\bigg\{\frac{1-\exp(-\tau/\lambda)}{\tau/\lambda}\bigg\}-\beta_2\exp\Big\{-\frac{\tau}{\lambda}\Big\}, \label{eqn_NS_model}
\end{equation}
where 
\begin{itemize}
\item $\beta_0$ is known as the long-run interest rate levels,
\item $\beta_1$ is the short-term effect,
\item $\beta_2$ is the midterm effect,
\item $\lambda$ is the decay factor.
\end{itemize}
The small value of $\lambda$ leads to slow decay and can better fit the curve at longer maturities.  Several literature \cite{Nelson_Siegel_1987, Diebold_Li_2006, Hays_Shen_Huang_2012, Chen_Niu_2014} reports that the model explains more than 90\% variations in yield curve. The movement of the parameters through time reflects the change in the monetory policy of Federal Reserve and hence the economic activity. The high corroleation indicates the ability of the fitted curves to predict the price of long term US Treasury bond. \emph{Estimations and statistical inference about the parameters are extremely important, as each parameter space} $\theta=(\beta_0, \beta_1, \beta_2,\lambda)$ \emph{of the model} (\ref{eqn_NS_model}), \emph{has its own economic interpretation}. In this chapter, we discuss a Bayesian approach for the estimation of the yield curve and further inference. 

\section{Why Bayesian Method?}

The Bayesian methods provides a consistent way of combining the prior information with data, within the decision theoretical framework. We can include past information about a parameter or hypothesis and form a prior distribution for the future analysis. When new observations become available, the previous posterior distribution can be used as the prior distribution. This inferences logically follow from the Bayes’ theorem. The Bayesian analysis presents inferences that are conditional on the data and are exact, without dependence on asymptotic approximation. When the sample size is small, the inference proceeds in the same way, as if one had a large sample. The Bayesian analysis can estimate any functions of parameters directly, without using the `plug-in' method.

In Bayesian inference, probability represents the degree of belief. In frequentist statistics, the probability means the relative frequency of an event. Therefore the frequentist method cannot assign the probability to a hypothesis (which is a belief), because a hypothesis is not an event characterized by a frequency. Instead, frequentist statistics can only calculate the probability of obtaining the data of an event, assuming a hypothesis is true. Therefore the Bayesian inference can calculate the probability that a hypothesis is valid, which is usually what the researchers want to know. By contrast, the frequentist statistics calculate the $p$-value, which is the probability of the more extreme data to obtain under the assumption that the null hypothesis is true. This probability, $\mathbb{P}(\text{data} | \text{null hypothesis is true})$, usually does not equal the probability that the null hypothesis is surely true. 

Frequentist statistics has only one well-defined hypothesis - the null hypothesis and alternate hypothesis is simply defined as the `null hypothesis' is wrong.  However, the Bayesian method can have multiple well-defined hypotheses. The frequentist methods transform the data into a test statistics and the $p$-value, then compares this value to an arbitrarily determined cutoff value and employs the decision-making approach to judge the significance, for example, reject the null hypothesis (or not reject) based on whether $p < \alpha,$ where $0 < \alpha <1$. The best way to do frequentist analysis would be to determine the sample size $n$ before you start collecting the  data. In Bayesian methods, because the probabilities represent the degrees of belief, it allows more nuanced and sophisticated analyses. We can calculate likelihoods and posterior probabilities for multiple hypotheses. We can enter data as we collect them; then update the degrees of belief so that we worry less about the arbitrary cut-off values. It makes sense to choose a hypothesis with maximum posterior probability, out of multiple hypotheses and not to worry about the arbitrary single value of significance. We can also do useful and straightforward analysis such as marginalizing over nuisance parameters, calculating likelihood ratio, or Bayes factor etc. In Bayesian methods, probability calculation follows the axiomatic foundation of the probability theory (e.g., the sum and product rules of the probability). By contrast, inferential frequentist method uses a collection of different test procedures that are not necessarily obtained from a coherent, consistent basis.

Having said that one should be aware of some possible disadvantages with the Bayesian methods. The prior distributions are often difficult to justify and can be a significant source of inaccuracy. There can be too many hypotheses, which may lead to the low posterior probability of each hypothesis, making the analysis sensitive to the choice of the prior distribution. The analytical solutions can be difficult to derive.  Analytical evaluation of posterior infernce can be intensive; but we can bypass this by using the state of the art Monte Carlo methods.

\section{Bayesian Approach to Modeling}

Bayesian approach to the statistical modeling follows three steps. First, we define the likelihood model, also known as the data model in some machine learning literature. In the second step, we describe the prior distributions, and the third step follows to obtain the posterior distribution model via Bayes theorem. Once we get the posterior model, all the Bayesian statistical inferences and predictions can be carried out based on the posterior model.

\subsection{Prior Distribution}
The prior probability distribution of an unknown parameter is the probability distribution that would express analysts beliefs about the quantity before any evidence is taken into consideration. We can develop a prior distribution, using a number of techniques \cite{Carlin_book_2008} describe below.

\begin{enumerate}
\item We can determine a prior distribution, from past data, if historical data exits.

\item We can elicite prior distribution, from the subjective assessment of an experienced expert in the domain. For example if an expert believe that long-term interest rate will never be more than 4\%, then that can be used to define the prior istribution for $\beta_0$. 

\item When no information is available, we can create an uninformative prior distribution to reflect a balance among outcomes.

\item  The prior distribution can also be chosen according to some objective principle, such as the maximizing entropy for given constraints. For examples:  the Jeffreys prior or Bernardo's reference prior. These priors are often known as the objective \item If the family of conjugate priors exists, then considering a prior from that family simplifies the further calculation.

\end{enumerate}

\subsubsection*{Prior for Interest rate levels}

In the Nelson-Siegel model as described in (\ref{eqn_NS_model}), the three interest rate parameters are: (i) $\beta_0$ is the long-run interest rate levels, (ii) $\beta_1$ represents short-term interest rate and (iii) $\beta_2$ represents medium-term interest rate. It is very rare that interest rate is negative. In fact, many argue if interest rate becomes negative then financial system collapse. So all practical purposes, we can assume that interest rates are positive and we can assume a prior probability distribution with its support only on the positive side of the real line. For example we can assume the inverse-gamma probability distribution over $\{\beta_0, \beta_1, \beta_2\}$. The fourth parameter of the model is the decay parameter $\lambda$, and it is natural to assign a prior distribution, the support of which is positive. Question is what would be a practical parameter value of the inverse gamma prior distribution? One choice could be the \texttt{Inverge-Gamma(a=1,b=1)}. The reason for such choice is if \texttt{a$\leq$1}, then the moments of the \texttt{Inverge-Gamma} distribution does not exist. If one does not have the idea about the mean and variance of the parameters, then such choice of prior could be used. Having said that one could check the $\mathbb{P}[0 \leq \beta \leq 30 ] \approx 0.97$, that is the prior belief; and there is 97\%  chance that interest rates are below 30\%. Such kind of probabilistic statement conveys a vague idea about the possible values of interest rate. We assumes that in the prior distributions the parameters are exchangebale and the there is no dependence among the parameters. So the first prior distribution we consider is
$$
\pi(\beta_0, \beta_1, \beta_2,\lambda,\sigma)=\pi(\beta_0)\pi(\beta_1)\pi(\beta_2)\pi(\lambda)\pi(\sigma),
$$
where $\beta_0, \beta_1, \beta_2,\lambda,\sigma \sim $ \texttt{Inverge-Gamma(a=1,b=1)}.

In the history of finance, the negative yield is though rare, it has occurred. Therefore it would be wise to consider an alternative model, which allows negative effect on yield. Therefore we consider \texttt{Normal}($\mu_{\beta},\sigma_{\beta}$) prior distribution over the interest rate and \texttt{Inverge-Gamma(a=1,b=1)} over the $\sigma_{\beta}$, which models the scaling effect of the interest rate. Such kind of model is also known as the hierarchical Bayes model. The detail of the second model presented as Model 2 in the table (\ref{tbl_prior_models}). Besides, we considered a slight variation of the Model 2 and named it as the Model 3. In the Model 3, we considered \texttt{Inverge-Gamma(a=0.1,b=0.1)} over the $\{\lambda,\sigma,\sigma_{\beta}\}$.

\subsection{Likelihood Function}

Now we discuss one of the most important concept of Statistics, known as the `likelihood'. Note that bothe frequestist and Bayesian statistics agrees that there has to be a data model or likelihood function to do any statistical inference. In order to understand the concept of the likelihood function, we consider a simple example.

\noindent \textbf{Example}

 Suppose $y$ is the number insurance claims that follow Poisson distribution 
$$
\mathbb{P}(Y=y)= e^{-\lambda}\frac{\lambda^y}{y!},~~ y=0,1,2,\hdots; 0<\lambda <\infty,
$$
and in the dataset we have only one data points and that is $Y=5$. We donot have any idea about $\lambda$. Different values of $\lambda$ will result into different values of $\mathbb{P}[Y=5]$.  Note that $\mathbb{P}[Y=5|\lambda=3]$ denotes the probability of $Y=5$, when $\lambda=3$. Here we compare the probabilities for different values of $\lambda$,
\begin{eqnarray*}
\mathbb{P}[Y=5|\lambda=3]&=&e^{-3}\frac{3^5}{5!}\approx 0.101,\\
\mathbb{P}[Y=5|\lambda=4]&=&e^{-4}\frac{4^5}{5!}\approx 0.156,\\
\mathbb{P}[Y=5|\lambda=5]&=&e^{-5}\frac{5^5}{5!}\approx 0.175,\\
\mathbb{P}[Y=5|\lambda=6]&=&e^{-6}\frac{6^5}{5!}\approx 0.161,\\
\mathbb{P}[Y=5|\lambda=7]&=&e^{-7}\frac{7^5}{5!}\approx 0.128.
\end{eqnarray*}
We can conclude that $\lambda=5$  justifies  the data almost 75\% better than $\lambda=3$.   That is because if the value of $\lambda$ is close to 5 then the the likelihood of the seeing the data `Y=5' is much higher than when $\lambda=3,4,6$ or $7$.

The model for the given data is presented as a function of the unknown parameter $\lambda$, is called likelihood function. The likelihood function can be presented as
$$
l(\lambda | Y=5)=p(5|\lambda).
$$
However, in reality we typically have the multiple observations like $\y=\{y_1,y_2,\hdots,y_n\}$. Then ofcourse we have to look into the joint probability models of $\y$. The likelihood function for such model would be
$$
l(\theta | y_1,y_2,\hdots,y_n)=p(Y_1=y_1,\hdots,Y_n=y_n|\theta),
$$
where $\theta$ is parameters of the model and $\theta$ could be scalar or vector, depending on the model.

\subsection*{Likelihood Function for Nelson-Siegel Model}
The  Nelson-Siegel function is believed to be the model which explain the behaviour of the yield curve rate. Now it is expected there will be some random shock or unexplained error in the observed rate. Hence the expected data model would be
\begin{eqnarray}\label{eqn_data_model}
y_i(\tau_j)=\mu_i(\tau_j)+e_{ij},
\end{eqnarray}
where 
$$
\mu_i(\tau_j)=\beta_0+(\beta_1+\beta_2)\bigg\{\frac{1-\exp(-\tau_j/\lambda)}{\tau_j/\lambda}\bigg\}-\beta_2\exp\Big\{-\frac{\tau_j}{\lambda}\Big\},
$$
$j=1,2,\hdots,m$, $i=1,2,\hdots,n$, and $e_{ij}\stackrel{i.i.d}{\sim} N(0,\sigma^2)$. Note that in the data model (\ref{eqn_data_model}), it is the error or unexplained part which is stochastic or random. The assumption of the \emph{independent and identically distributed} (aka. \emph{i.i.d})  error provides us to assume that each observations $y_{ij}$ independently follow
$$
y_i(\tau_j) \stackrel{indep}{\sim} N\big(\mu_i(\tau_j),\sigma^2\big).
$$ 
The likelihhod function of the Nelson-Siegel function can be modeled as
\begin{equation}\label{eqn_likelihhod_NS}
l(\mathcal{D}|\theta) = \prod_{i=1}^{n}\prod_{j=1}^{m}p\big(\mu_i(\tau_j),\sigma^2\big),
\end{equation}
where $\theta=(\beta_0,\beta_1,\beta_2,\lambda,\sigma^2)$ is the parameter vector needs to be estimated and $p(.)$ is the probability density function (pdf) of the Gaussian distribution, $\mathcal{D}=\{y_{11},\hdots,y_{nm},\tau_1,\hdots,\tau_m\}$ is the data or evidence.

\subsection{Posterior Distribution}

The posterior probability distribution of unknown parameters, conditional on the data obtained from an experiment or survey.  The ``Posterior,'' in this context, means after taking into account the relevant data related to the particular study. The posterior probability of the parameters $\theta$ given the data $\mathcal{D}$ is denoted as $p(\theta|\mathcal{D})$. On the contrary, the likelihood function is the probability of the evidence given the parameters is denoted as the $l(\mathcal{D}|\theta)$. The concepts are related via Bayes theorem as
\begin{eqnarray}
p(\theta|\mathcal{D})&=&\frac{l(\mathcal{D}|\theta)~p(\theta)}{\int_{\Theta}l(\mathcal{D}|\theta)~p(\theta)~d\theta} \nonumber \\
&=&\frac{l(\mathcal{D}|\theta)~p(\theta)}{p(\mathcal{D})}.\label{eqn_posterior}
\end{eqnarray}
There are two points to note.
\begin{itemize}
  \item The denominator of the (\ref{eqn_posterior}), is free of $\theta$. Therefore, often the posterior model is presented as proportion of likelihood times prior, i.e.,
$$
p(\theta|\mathcal{D}) \propto l(\mathcal{D}|\theta)~p(\theta).
$$
  \item The integration of the in the denominator of (\ref{eqn_posterior}) is high-dimensional integration problem. For example, in the Model 2 of the table (\ref{tbl_prior_models}), there are six parameters in the $\theta$, i.e., $\theta=\{\beta_0,\beta_1,\beta_2,\lambda,\sigma,\sigma_{\beta}\}$. So the integration will be a six-dimensional integration  problem.
  \item Since having an analytical solution of the six-dimensional integration problem is almost impossible; we resort to Monte Carlo simulation methods.
\end{itemize}

\subsection{Posterior Inference}

In Bayesian methodology, the posterior model for the parameter $\theta$, contains all the information. However, that is too much information to process. Hence we look for the  summary statistics of the posterior probability distribution. The Bayesian estimation methods consider the  measurements of central tendency of the posterior distribution as the representative value of the parameter. The measures are:
\begin{itemize}
\item \textbf{Posterior Median}: For one-dimensional problems, an unique median exists for the real valued parameters. The posterior median is also known as the robust estimator. If $\int_{\mathbb{R}}p(\theta~|~\mathcal{D}) < \infty$, then posterior median  $\tilde{\theta}$ is
\begin{equation}\label{post_median}
\mathbb{P}(\theta \leq \tilde{\theta}~|~\mathcal{D}~) =\int_{-\infty}^{\tilde{\theta}}p(\theta~|~\mathcal{D})~d\theta = \frac{1}{2}.
\end{equation}
The posterior median is Bayes estimator under squared error loss function \cite{berger1985statistical}.
\item \textbf{Posterior Mean}:  If there exists a finite mean for the posterior distribution, then we can consider the posterior mean as the estimate of the parameter, i.e.,
\begin{equation}\label{post_mean}
\hat{\theta}= \mathbb{E}(\theta~|~\mathcal{D}~) = \int_{\Theta}\theta p(\theta~|~\mathcal{D})~ d \theta.
\end{equation}
The posterior mean is Bayes estimator under squared error loss function \cite{berger1985statistical}.
\item \textbf{Posterior Mode}: The mode of the posterior distribution, also known as the  maximum a posteriori probability (MAP) estimate, 
\begin{equation}\label{post_mode}
\bar{\theta} =\argmax_{\Theta} p(\theta~|~\mathcal{D}~).
\end{equation}
The posterior mode is Bayes estimator under Kullback-Leibler type loss
function \cite{Das.Dey.2010}.
\end{itemize}
Note that the posterior mean and the posterior median is an integration problem and the posterior mode is an optimization problem. 

\subsection{Posterior Analysis of US Treasury Yield Data}

Here we present the Bayesian posterior analysis of the Nelson-Siegel model (\ref{eqn_NS_model}), with three different prior distribution models,  presented in the table (\ref{tbl_prior_models}). We worked all computational issues using the \texttt{rstan} package in \texttt{R} statistical software. In this analysis, considered only six days of data (from May 01, 2018 to May 08, 2018) presented in the Table \ref{tbl_US_treasury_data}. \emph{Note that the purpose of this toy analysis is to demonstrate how the Bayesian analysis works}!

\begin{figure}[h!]
\centering
\includegraphics[height=3in, width=7in]{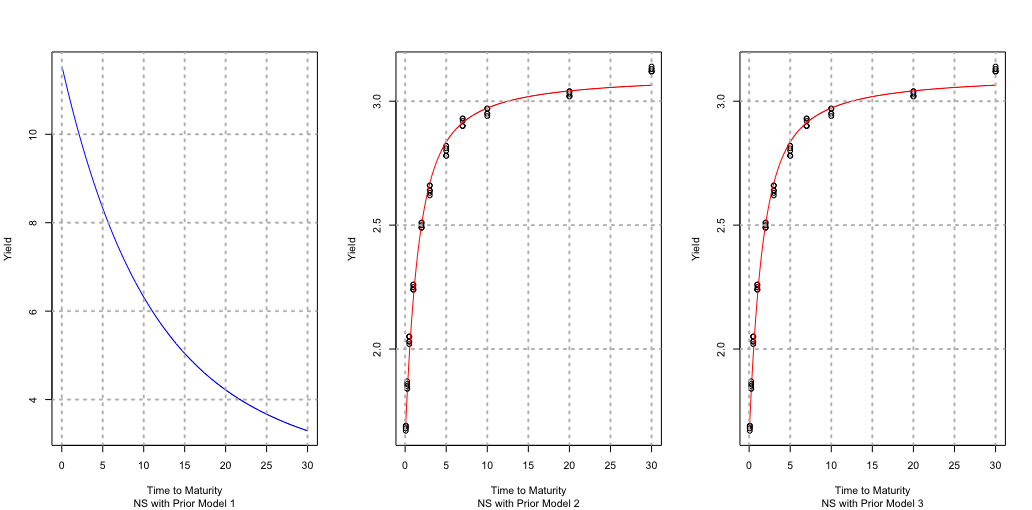}
\caption{Fitted Nelson Siegel yield curve with the three prior models described in table \ref{tbl_prior_models}. The fitted yield curve with the prior model 1 is unrealistic, indicates undesirable prior distribution. The fitted yield curve with prior model 2 and 3 show an excellent fit to data.}\label{fig_NS_fit}
\end{figure}

The figure \ref{fig_NS_fit}, exhibit the fitted Nelson Siegel yield curve with the three different prior distributions illustrated in table \ref{tbl_prior_models}. The fitted yield curve with the prior model 1 is unrealistic, indicates an unsatisfactory prior distribution. The fitted yield curve with prior model 2 and 3 show an excellent fit to data. The MAP estimates for the Nelson-Siegel parameters is presented in the table \ref{tbl_map_estimates}. , with the three prior distributions in table \ref{tbl_prior_models}. In case of the model 1, the long-run effect $\beta_0$ for the prior model 1, is less than the medium-term effect $\beta_2$, makes the prior model 1 an undesirable. The similar MAP estimates of the Nelson-Siegel parameters for model 2 and 3 indicates robust posterior analysis, in spite of differences in the prior parameters. 

As we see the model 1 is really undesirable, hence we drop this model from the further discussion and we only focus our discussion on the hierarchical model considered in model 2 and 3. \textbf{Having said that one must note that the Bayesian methodology is not a magic bullet. A poorly chosen prior distribution may lead to an undesirable model}. The Monte Carlo estimates of the posterior mean, standard deviation, median, 2.5\% and 97\% quantile of the Nelson-Siegel parameters under the prior distribution model 2 and 3, under table \ref{tbl_mc_estimate}. 

\begin{figure}[ht]
\centering
\includegraphics[height=8in, width=7in]{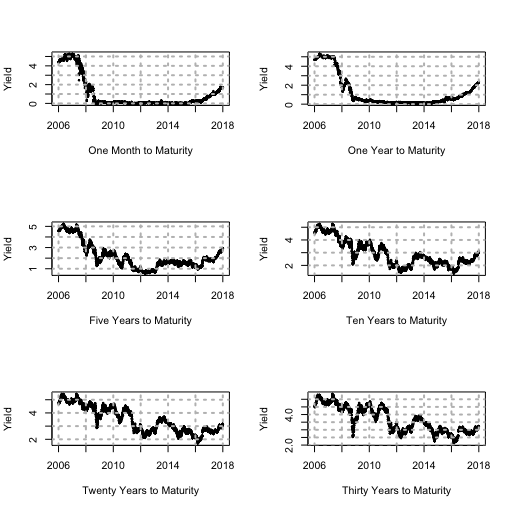}
\caption{US Tresury's yield rate (from 02-Oct-2006 to 08-May-2018) presented in six panels. Subtitle of each panel state the yield rate corresponding to the `time to maturity'. The x-axis present the years and y-axis represent the yield rate.}\label{fig_UST_yield}
\end{figure}

\begin{figure}[ht]
\centering
\includegraphics[height=3in, width=7in]{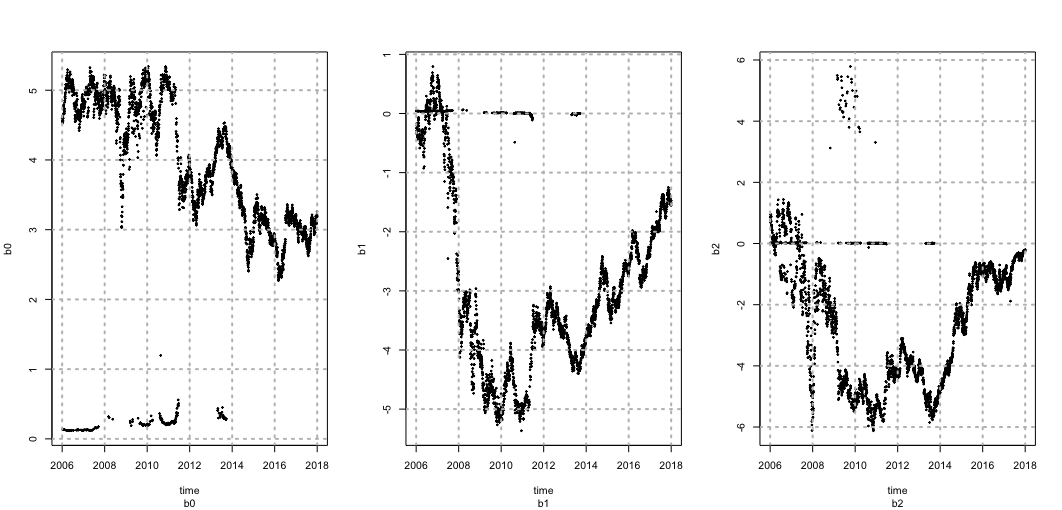}
\caption{The MAP estimates of the Nelson-Siegel Parameters from the US Treasury Yield data \cite{US_treasury_sample_data} and presented in the figure \ref{fig_UST_yield}.}\label{fig_NSParameters_2006_2018}
\end{figure}

\begin{figure}[ht]
\centering
\includegraphics[height=4in, width=4in]{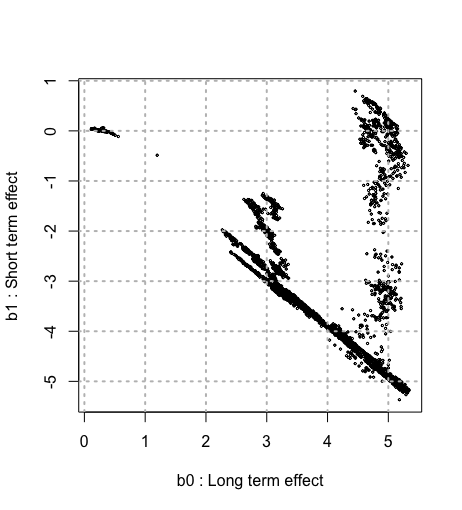}
\caption{Scatter plot of daily MAP values of $\beta_0$ and $\beta_1$ of Nelson-Siegel Model.  The plot indecates a negative relation between the two parameters. However, we assume independe among all the parameters in the prior distribution.}\label{fig_NSParameters_relation}
\end{figure}

We present the US Treasury yield curve data in the figure \ref{fig_UST_yield}, and present the MAP estimate of the Nelson-Siegel parameters in the figure \ref{fig_NSParameters_2006_2018}. We present the scatter plot of the daily MAP values of $\beta_0$ and $\beta_1$ of Nelson-Siegel Model in the figure \ref{fig_NSParameters_relation}.  The plot indicates a negative relationship between the long and short-term effect. However, we assume independence among all the parameters in the prior distribution.

\section{Dynamic Nelson-Siegel Model}

The Dynamic Nelson-Siegel (DNS) model \citep{Nelson_Siegel_1987, Diebold_Li_2006,Chen_Niu_2014} for yield curve can be presented as 
\begin{eqnarray*}
y_t(\tau_j)&=&\beta_{1t}+\beta_{2t}\bigg(\frac{1-\exp\{-\tau_j/\lambda\}}{\tau_j/\lambda}\bigg)+\beta_{3t}\bigg(\frac{1-\exp\{-\tau_j/\lambda\}}{\tau_j/\lambda}-\exp\{-\tau_j/\lambda\}\bigg)+\epsilon_t(\tau_j),\\
\epsilon_t(\tau_j)&\sim& N(0,\sigma_{\epsilon}^2),\\
\beta_{it}&=&\theta_{0i}+\theta_{1i}\beta_{i,t-1}+\eta_{i},~~i=1,2,3,~~\eta_i\sim N(0,\sigma_{\eta}^2),~~t=1,2,\hdots, T,~~~j=1,2,\hdots,m,
\end{eqnarray*}
where $y_t(\tau)$ is the yield for maturity $\tau$ (in months) at time $t$. The three factors $\beta_{1t}$, $\beta_{2t}$ and $\beta_{3t}$ are denoted as level, slope and curvature of slope respectively. Parameter $\lambda$ controls exponentially decaying rate of the loadings for the slope and curvature. The goodness-of-fit of the yield curve is not very sensitive to the specific choice of $\lambda$ \citep{Nelson_Siegel_1987}. Therefore \cite{Chen_Niu_2014} chose $\lambda$ to be known. In practice, $\lambda$ can be determined through grid-search method. There are eight static parameters $\bft=(\theta_{01},\theta_{02},\theta_{03},\theta_{11},\theta_{12},\theta_{13},\sigma_{\epsilon}^2,\sigma_{\eta}^2)$ in the model. In matrix notation the DNS model can be presented as 
\begin{eqnarray}
\bfb_{t}&=&\theta_0+\Z\bfb_{t-1}+\bfeta_t, \label{system_equation}\\
\y_t&=&\bm{\phi}\bfb_t+\bfe_t,\label{observation_equation}
\end{eqnarray}
where
$\y_t=\left(\begin{array}{c}
y_t(\tau_1)\\
y_t(\tau_2)\\
\vdots\\
y_t(\tau_m)
\end{array}
\right)_{m \times 1}$,
$\bm{\phi}=\left(\begin{array}{ccc}
1& f_1(\tau_1) & f_2(\tau_1)\\
1& f_1(\tau_2) & f_2(\tau_2)\\
\vdots & \vdots & \vdots \\
1& f_1(\tau_m) & f_2(\tau_m)\\
\end{array}
\right)_{m \times 3}$,
$
\bfb_t=\left(\begin{array}{c}
\beta_{0t}\\
\beta_{1t}\\
\beta_{2t}
\end{array}
\right)_{3 \times 1},
$
$\bfe_t=\left(\begin{array}{c}
\epsilon_1\\
\epsilon_2\\
\vdots\\
\epsilon_m
\end{array}
\right)_{m \times 1}$,
such that $f_1(\tau_j)=\big(\frac{1-\exp\{-\tau_j/\lambda\}}{\tau_j/\lambda}\big)$ and $f_2(\tau_j)=\big(\frac{1-\exp\{-\tau_j/\lambda\}}{\tau_j/\lambda}-\exp\{-\tau_j/\lambda\}\big)$, $j = 1,2,...,m$,
$\theta_0=\left(\begin{array}{c}
\theta_{01}\\
\theta_{02}\\
\theta_{03}
\end{array}
\right)$
and $\Z=\left(\begin{array}{ccc}
\theta_{11}&0&0\\
0&\theta_{12}&0\\
0&0&\theta_{13}
\end{array}
\right)$.
Note that $\bfe_t \sim \bm{N}_m(0,\sigma_{\epsilon}^2\bm{I}_m)$ and $\bfeta_t \sim \bm{N}_3(0,\sigma_{\eta}^2\bm{I}_3)$. Note that (\ref{system_equation}) is \textit{system equation} and (\ref{observation_equation}) is \textit{observation equation}. If available, we can use the generalized linear models (GLM) to incorporate any additional predictor variable, see \cite{Das2013,Das2006}.

\subsection{Relation between DNS model and Kalman Filter}\label{Relation_DNS_KF}
The term ``Kalman filter" refers to recursive procedure for inference. A beautiful tuorial paper on the same was wrtten by \cite{MeinholdSingpurwala1983}. The key notion here is that given the data $\Y_t = (\y_t,\y_{t-1},\hdots,\y_1)$ inference about $\bfb_{t}$ and prediction about $\y_{t+1}$ can be carried via Bayes theorem, which can be expressed as
\begin{eqnarray}
\mathbb{P}(\bfb_{t}|\Y_t)\propto \mathbb{P}(\y_t|\bfb_{t},\Y_{t-1})\times \mathbb{P}(\bfb_{t}|\Y_{t-1}).\label{Bayes_theorem}
\end{eqnarray}
Note that the expression on the left of equation (\ref{Bayes_theorem}) is the \textit{posterior distribution} of $\bfb$ at time $t$, whereas the first and second expression on the left side of (\ref{Bayes_theorem}) is the \textit{likelihood} and \textit{prior distribution} of $\bfb$, respectively. At $t-1$, our knowledge about $\bfb_{t-1}$ is incorporated in the probability statement for $\bfb_{t-1}$:
\begin{eqnarray}
(\bfb_{t-1}|\Y_{t-1})\sim \bm{N}_3(\hat{\bfb}_{t-1},\Sigma_{t-1}),\label{posterior_beta_at_t-1}
\end{eqnarray}
where $\hat{\bfb}_{t-1}$ and $\Sigma_{t-1}$ are the expectation and the variance of $(\bfb_{t-1}|\Y_{t-1})$. In effect, (\ref{posterior_beta_at_t-1}) is the posterior distribution of $\bfb_{t-1}$. We now look forward to time $t$ in two steps.
\begin{enumerate}
\item prior to observing $\y_t$,  
\item posterior or after observing $\y_t$, and
\item inference about $\y_{t}^*$ at maturity $\tau^*$.
\end{enumerate}
\noindent \textbf{\textit{Step 1}}: Prior to observing $\y_t$, our best choice for $\bfb_{t}$ is governed by the system equation (\ref{system_equation}) and is given as $\theta_0+\Z\bfb_{t-1}+\bfeta_t$. Since $\bfb_{t-1}$ is describe in (\ref{posterior_beta_at_t-1}), therefore
\begin{eqnarray}
(\bfb_{t}|Y_{t-1})\sim \bm{N}_3(\theta_0+\Z\hat{\bfb}_{t-1}~,~R_t=\Z\Sigma_{t-1}\Z^T+\sigma_{\eta}^2\bm{I}_3)\label{prior_beta_at_t}
\end{eqnarray}
is the prior distribution of $\bfb$ at time $t$. In obtaining (\ref{prior_beta_at_t}) we use the result for any constant $B$,
$$
X\sim \bm{N}(\mu,\Sigma)\implies a+BX \sim \bm{N}(a+B\mu,B\Sigma B^T).
$$
\noindent \textbf{\textit{Step 2}}: On observing $\y_t$, our objective is to obtain the posterior $\bfb_t$ using (\ref{Bayes_theorem}). However, to do this, we need the likelihood $\mathcal{L}(\bfb_t|\Y_t)$, or equivalently $\mathbb{P}(\y_t|\bfb_{t},\Y_{t-1})$. Let $e_t$ is the error in predicting $\y_t$ from previous time point $t-1$; thus
\begin{eqnarray}
\e_t=\y_t-\hat{\y}_t=\y_t-\bm{\phi}\theta_0-\bm{\phi}\Z\hat{\bfb}_{t-1}.\label{error_model}
\end{eqnarray}
Since, $\bm{\phi}$, $\Z$, $\theta_0$ and $\hat{\bfb}_{t-1}$ are known, observing $\y_t$ is equivalent to observing $\e_t$. Therefore (\ref{Bayes_theorem}) can be expressed as:
\begin{eqnarray*}
\mathbb{P}(\bfb_{t}|\y_t,\Y_{t-1})=\mathbb{P}(\bfb_{t}|\e_t,\Y_{t-1})\propto \mathbb{P}(\e_t|\bfb_{t},\Y_{t-1})\times \mathbb{P}(\bfb_{t}|\Y_{t-1}),
\end{eqnarray*}
where $\mathbb{P}(\e_t|\bfb_{t},\Y_{t-1})$ is the likelihood. Using the fact that $\y_t=\bm{\phi}\bfb_t+\bfe_t$, (\ref{error_model}) can be expressed as $\e_t=\bm{\phi}(\bfb_t-\theta_0-\Z\hat{\bfb}_{t-1})+\bfe_t$, so that 
$
\mathbb{E}(\e_t|\bfb_t,\Y_{t-1})=\bm{\phi}(\bfb_t-\theta_0-\Z\hat{\bfb}_{t-1}).
$
Since, $\bfe_t\sim \bm{N}_m(0,\sigma_{\epsilon}^2\bm{I}_m)$, it follows the likelihood as
\begin{eqnarray}
(\e_t~|~\bfb_t,\Y_{t-1})\sim \bm{N}_m(\bm{\phi}(\bfb_t-\theta_0-\Z\hat{\bfb}_{t-1})~,~\sigma_{\epsilon}^2\bm{I}_m).\label{likelihood_error}
\end{eqnarray}
Now in order to find the posterior, we use the standard result of the Gaussian distribution (\cite{Anderson1984}, pp. 28--30 ). If $X_1\sim N(\mu_1,\Sigma_{11})$ and
\begin{eqnarray}
(X_2|X_1=x_1)\sim N(\mu_2+\Sigma_{21}\Sigma_{11}^{-1}(x_1-\mu_1)~,~\Sigma_{22}-\Sigma_{21}\Sigma_{11}^{-1}\Sigma_{12}),\label{conditional_dist}
\end{eqnarray}
 then
\begin{eqnarray}
\left(\begin{array}{c}
X_1\\
X_2
\end{array}
\right)\sim \bm{N}
\left[
\bigg(
\begin{array}{c}
\mu_1\\
\mu_2
\end{array}
\bigg)
,
\bigg(
\begin{array}{cc}
\Sigma_{11} & \Sigma_{12}\\
\Sigma_{21} & \Sigma_{22}\\
\end{array}
\bigg)
\right].\label{multivariate_normal_dist}
\end{eqnarray}
In our case,  lets consider $X_1\iff \bfb_t$ and $X_2 \iff \e_t$. Since $(\bfb_{t}|Y_{t-1})\sim \bm{N}_3(\theta_0+\Z\hat{\bfb}_{t-1}~,~R_t)$, we note that 
$$
\mu_1 \iff \theta_0+\Z\hat{\bfb}_{t-1}~~ \text{and} ~~ \Sigma_{11} \iff R_t.
$$
If in (\ref{conditional_dist}), we replace $X_1$,$X_2$, $\mu_1$ and $\Sigma_{11}$ by $\bfb_t$, $\e_t$, $\theta_0+\Z\hat{\bfb}_{t-1}$ and $R_t$ respectively and compare the result (\ref{likelihood_error}), then
$$
\mu_2+\Sigma_{21}R_t^{-1}(\bfb_t-\theta_0-\Z\hat{\bfb}_{t-1})\iff \bm{\phi}(\bfb_t-\theta_0-\Z\hat{\bfb}_{t-1}),
$$
so that $\mu_2\iff \bm{0}$ and $\Sigma_{21} \iff \bm{\phi} R_t$; following the same method
$$
\Sigma_{22}-\Sigma_{21}\Sigma_{11}^{-1}\Sigma_{12}= \Sigma_{22}- \bm{\phi} R_t \bm{\phi}^T \iff \sigma_{\epsilon}^2\bm{I}_m,
$$
so that $\Sigma_{22} \iff \bm{\phi} R_t \bm{\phi}^T + \sigma_{\epsilon}^2\bm{I}_m$. Under the result (\ref{conditional_dist}) and (\ref{multivariate_normal_dist}) the joint distribution of $\bfb_t$ and $\e_t$, given $\Y_{t-1}$ can be described as
\begin{eqnarray*}
\left(\begin{array}{c}
\bfb_t\\
\e_t
\end{array}
\bigg | \Y_{t-1}\right)\sim \bm{N}
\left[
\bigg(
\begin{array}{c}
\theta_0+\Z\hat{\bfb}_{t-1}\\
\bm{0}
\end{array}
\bigg)
,
\bigg(
\begin{array}{cc}
R_t~ &~ R_t^T\bm{\phi}^T \\
\bm{\phi} R_t~ & ~\bm{\phi} R_t \bm{\phi}^T + \sigma_{\epsilon}^2\bm{I}_m\\
\end{array}
\bigg)
\right].
\end{eqnarray*}
So we have the posterior distribution of $\bfb_t$ at time point $t$ is
$$
(\bfb_t|\Y_{t})=(\bfb_t|\e_t,\Y_{t-1})\sim \bm{N}(\hat{\bfb}_t,\Sigma_t),
$$
where
\begin{eqnarray}
\hat{\bfb}_t=\mathbb{E}(\bfb_t|\Y_{t})=\theta_0+\Z\hat{\bfb}_{t-1}+R_t^T\bm{\phi}^T[\bm{\phi} R_t \bm{\phi}^T + \sigma_{\epsilon}^2\bm{I}_m]^{-1}\e_t,\label{posterior_mean_of_beta}
\end{eqnarray}
and
$$
\Sigma_{t}=R_t-R_t\bm{\phi}^T[\bm{\phi} R_t \bm{\phi}^T + \sigma_{\epsilon}^2\bm{I}_m]^{-1}\bm{\phi} R_t.
$$
\noindent \textbf{\textit{Step 3}}: Now in order to predict yield at a new maturity point(s) $\tau^*$ we can simply plug-in $\hat{\bfb}_t$ in observation equation (\ref{observation_equation}), i.e., 
\begin{eqnarray}
\hat{y}_t(\tau^*)=\bm{\phi}(\tau^*)\hat{\bfb}_t.\label{DNS_yield_predict}
\end{eqnarray}
\subsection{Gaussian Process Prior}
The Gaussian process prior for DNS is presented by \cite{sambasivan2017statistical}. This can be accomplished very easily by introducing a random component in observation equation (\ref{observation_equation}). The modeified observation equation is
\begin{eqnarray*}
\y_t &=&\bm{\phi}\bfb_t+W_t(\tau)+\bfe_t,\label{modified_observation_equation}
\end{eqnarray*}
where $\y_t$, $\bm{\phi}$, $\bfb_t$ and $\bfe_t$ are defined as in (\ref{observation_equation}) and $W_t(\tau) \sim \bm{N}_m(\bm{0},\mathbf{K})$, where $\mathbf{K}=\rho(\tau,\tau')$. Following the structure of the GP model (\cite{Rasmussen2005}), at time point $t$ is
\begin{eqnarray}
f_t&\sim& \bm{N}_m\left(\bm{\phi}\bfb_t,\mathbf{K}\right), ~~\bfe_t \sim \bm{N}_m(0,\sigma_{\epsilon}^2\bm{I}_m)\nonumber \\ 
\y_t &\sim& \bm{N}_m(\bm{\phi}\bfb_t,\mathbf{K}+\sigma_{\epsilon}^2\bm{I}_m).\label{likelihood_GP}
\end{eqnarray}

We consider the same system equation as (\ref{system_equation}). Since the system equation is same, therefore step 1 and 2 for DNS with the Gaussian process prior would be same as in the section \ref{Relation_DNS_KF}. 

\subsection{Marginal Likelihood}

It will be useful to compute the probability that DNS with a given set of parameters (prior distribution, transition and observation models) would produce an observed signal. This probability is known as the `marginal likelihood' because it integrates out the hidden state variables $\bfb_t$, so it can be computed using only the observed data $\y_t$. The marginal likelihood is useful to estimate different static parameter choices using Bayesian computation technique.

It is easy to estimate the marginal likelihood as a side effect of the recursive filtering calculation. By the chain rule, the likelihood can be factored as the product of the probability of each observation given previous observations,
\begin{eqnarray*}
p(\Y|\bft)=\prod_{t=0}^T p(\y_t|\y_{t-1},\y_{t-2},\hdots,\y_0,\bft)
\end{eqnarray*}
and because the Kalman filter describes a Markov process, all relevant information from previous observations is contained in the current state $(\bfb_t|\Y_{t-1})$. Note that $\bft$ is the static parameter(s). Thus the marginal likelihood is given by 
\begin{eqnarray*}
p(\Y_T|\bft)&=&\prod_{t=0}^T p(\y_t|\Y_{t-1},\bft)d\y_t\\
&=&\prod_{t=0}^T \int p(\y_t|\bfb_t)p(\bfb_t|\Y_{t-1})d\bfb_t\\
&& \text{ consider the likelihood (\ref{likelihood_GP}) and prior at time $t$ (\ref{prior_beta_at_t})}\\
&=& \prod_{t=0}^T \int \bm{N}_m(\y_t;\bm{\phi}\bfb_t,\tilde{\mathbf{K}})\bm{N}_3(\bfb_t;\hat{\bfb}_{t|t-1}~,~R_t)d\bfb_t\\
&& \text{where  } \tilde{\mathbf{K}}=\mathbf{K}+\sigma_{\epsilon}^2\bm{I}_m \text{ and } \hat{\bfb}_{t|t-1}=\theta_0+\Z\hat{\bfb}_{t-1}\\
&=& \prod_{t=0}^T \bm{N}_m(\y_t;\bm{\phi}\hat{\bfb}_{t|t-1},\tilde{\mathbf{K}}+ \bm{\phi}R_t\bm{\phi}^T),\\
&=& \prod_{t=0}^T \bm{N}_m(\y_t;\bm{\phi}\hat{\bfb}_{t|t-1},\bm{S}_t), \text{ where }\bm{S}_t=\tilde{\mathbf{K}}+ \bm{\phi}R_t\bm{\phi}^T
\end{eqnarray*}
i.e., product of multivariate normal densities. This can easily be calculated as a simple recursive update. However, to avoid numeric underflow, it is usually desirable to estimate the log marginal likelihood $l=\log p(\Y_T|\bft)$. We can do it via recursive update
$$
l^{(t)}=l^{(t-1)}-\frac{1}{2}\Big\{\ln|\bm{S}_t|+ m\ln 2\pi+(\y_t-\bm{\phi}\hat{\bfb}_{t|t-1})\bm{S}_t^{-1}(\y_t-\bm{\phi}\hat{\bfb}_{t|t-1})^T\Big\}.
$$
Note that computation of $\bm{S}_t^{-1}$ involves the complexity of $O(n^{3})$, where $n$ is the number of data point. For fast GP regression, see \cite{sambasivan2018fastGP}.

\section{Computational Issues}

In general, it is impossible to obtain explicit analytical form for MAP (\ref{post_mode}), posterior mean (\ref{post_mean}), or posterior median (\ref{post_median}). This implies that we have to resort to numerical methods, such as the Monte Carlo, or optimization subroutine. We implement the optimization for MAP estimates using the BFGS method.  This
method has the time complexity of $O(p^{2})$ per iteration where $p$ is the
number of parameters. The order of convergence for BFGS method is super-linear.

We implement the posterior mean, posterior median, via the Hamiltonian Monte Carlo (HMC) algorithm for hierarchical models \cite{Betancourt.2013, Hoffman.Gelman.2014}, using the \texttt{rstan} software \cite{stan_manual}. The \texttt{rstan} can also implement the BFGS optimization method.

\subsection{Monte Carlo Pricing of Bond}

An interesting by-product of the Monte Carlo method is it helps us to estimate the theoretical price of a bond. We know the price of a bond is a non-linear function of the yield curve. That is
\begin{equation}\label{eqn_bond_pricing}
P=f\big(Y(\tau,\theta)\big),
\end{equation}
where $P$ is the price of the bond and $Y(\tau)$ is the yield curve modeled by the Nelson-Siegel function (\ref{eqn_NS_model}). Suppose $\{\theta_1^{*}, \theta_2^{*},\hdots, \theta_M^{*}\}$ are the Monte Carlo simulation of the $\theta$ in (\ref{eqn_NS_model}). Then we can plug-in each of the $\theta_i^*$ in the pricing equation (\ref{eqn_bond_pricing}) and we can get the Monte Carlo price $\{P_i^*~|~i=1,2,\hdots,M\}$ , where $M$ is the simulation size. Now we can estimate the posterior mean, median and $100\times (1-\alpha)\%$ confidence interval for the price of the bond. If $P_l$ is the lower bound and $P_u$ is the upper bound of the interval, and if the `traded price' is below the $P_l$ then that will indicate that the bond is undervalued. Similarly, if the `traded price' is above the $P_u$, then that will indicate that the bond is overvalued.

\vspace{0.5in}

\noindent \textbf{Experiment}: We demonstrate the concept with a simple experiment. Suppose we have a bond which will mature in 15 years, pays coupon twice a year at 4\% annual rate with a par value of \$ 1000. If we have yield data as presented in table \ref{tbl_US_treasury_data}; what would be the Bayesian price of the bond on May 9th, 2018?

\begin{figure}[ht]
\centering
\includegraphics[height=4in, width=4in]{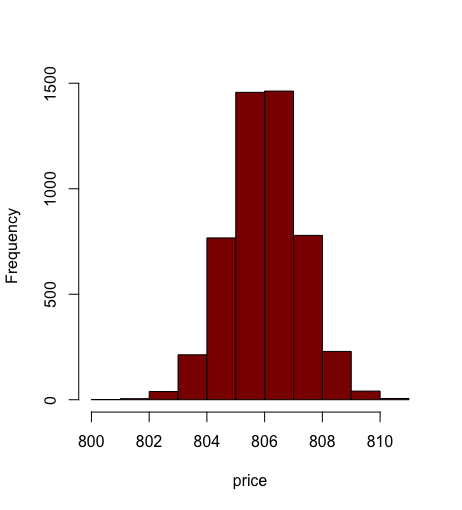}
\caption{Histogram of 5000 Monte Carlo price of the bond.}\label{fig_MC_price_hist}
\end{figure}

\begin{figure}[ht]
\centering
\includegraphics[height=4in, width=4in]{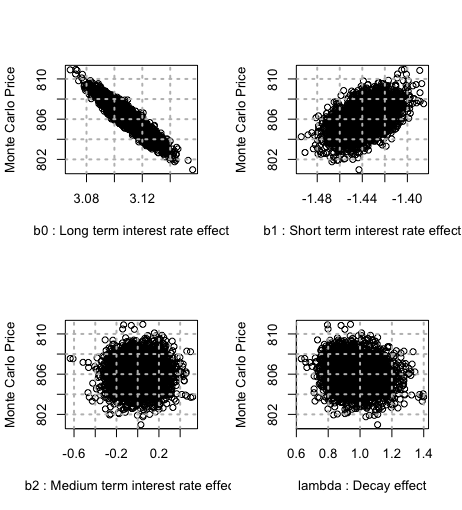}
\caption{Relatioship between the Monte Carlo price of the bond and the parameters of the Nelson-Siegel Model. We can see the strong negative correlation between the price and long-term effect of yield, i.e., $\beta_0$ and a weak positive correlation between short-term interest rate effect $\beta_1$ and the value of the bond.}\label{fig_MC_price_vs_NSParam}
\end{figure}

We considered the prior model 3 presented in the table \ref{tbl_prior_models} for this task. The parameter values of the Nelson-Siegel model were simulated from the posterior model using the No-U turn sampler of the HMC algorithm via \texttt{rstan} package. Then we calculate the 5000 Monte Carlo price of the bond and present histogram of the 5000 simulated price in the figure \ref{fig_MC_price_hist}. We present the posterior summary of the bond price in the table \ref{tabl_posterior_summary}. The expected price is \$ 806.01, and the traded price should stay within (\$ 803.59, \$ 808.46). If the traded price goes below \$ 803.59, then we can consider the bond to be undervalued; while if the traded price goes above \$ 808.46 then we can consider the bond to be over-priced. The figure \ref{fig_MC_price_vs_NSParam} presents an exciting relationship between the Monte Carlo price of the bond and parameters of the Nelson-Siegel function. We can see the strong negative correlation between the price and long-term effect of yield, i.e., $\beta_0$ and a weak positive correlation between short-term interest rate effect and the value of the bond.

\section{Conclusion}

In this work, we present the hierarchical Bayesian methodology to model the Nelson-Siegel yield curve model. We demonstrate that ad-hoc choice of prior may lead to undesirable results. However, the proposed the hierarchical Bayesian method is much more robust and deliver the desired effect. We used BFGS algorithm in \texttt{rstan} for the MAP estimates of the Nelson-Siegel's parameters. We also implemented full Bayesian analysis using the HMC algorithm available in \texttt{rstan} package. As a by-product of the HMC, we simulate the Monte Carlo price of a Bond, and it helps us to identify if the bond is over-valued or under-valued. We demonstrate the process with example and US treasury's yield curve data. One interesting finding is that there is a strong negative correlation between the price and long-term effect of yield, i.e., $\beta_0$. However, the relationship between the short-term interest rate effect and the value of the bond is weakly positive. This is phenomenon is observed because the posterior analysis shows an inverse relationship between the long-term and the short-term effect of the Nelson-Siegel model.


\bibliographystyle{authortitle}
\bibliography{sample}

\begin{thebibliography}{10}

\bibitem{US_treasury_sample_data}
Us treasury yield curve rates.
\newblock
  \url{https://www.treasury.gov/resource-center/data-chart-center/interest-rates/Pages/TextView.aspx?data=yield}.
\newblock Accessed: 2018-May-08.

\bibitem{Anderson1984}
T.~W. Anderson.
\newblock {\em An Introduction to Multivariate Statistical Analysis}.
\newblock Wiley, 1984.

\bibitem{berger1985statistical}
James~O. Berger.
\newblock {\em Statistical decision theory and Bayesian analysis}.
\newblock Springer-Verlag, New York, 1985.

\bibitem{Betancourt.2013}
M.~Betancourt and M.~Girolami.
\newblock Hamiltonian monte carlo for hierarchical models.
\newblock Technical report, 2013.

\bibitem{Carlin_book_2008}
Bradley~P Carlin and Thomas~A. Louis.
\newblock {\em Bayesian Methods for Data Analysis}.
\newblock CRC Press., third edition, 2008.

\bibitem{Chen_Niu_2014}
Ying Chen and Linlin Niu.
\newblock Adaptive dynamic nelson-siegel term structure model with
  applications.
\newblock {\em Journal of Econometrics}, 180(1):98--115, 2014.

\bibitem{Das2006}
Sourish Das and Dipak Dey.
\newblock On bayesian analysis of generalized linear models using jacobian
  technique.
\newblock {\em The American Statistician}, 60, 2006.

\bibitem{Das2013}
Sourish Das and Dipak Dey.
\newblock On dynamic generalized linear models with applications.
\newblock {\em Methodology and Computing in Applied Probability}, 15, 2013.

\bibitem{Das.Dey.2010}
Sourish. Das and Dipak.~K Dey.
\newblock On bayesian inference for generalized multivariate gamma
  distribution.
\newblock {\em Statistics and Probability Letters}, 80:1492--1499, 2010.

\bibitem{Diebold_Li_2006}
Francis Diebold and Canlin Li.
\newblock Forecasting the term structure of government bond yields.
\newblock {\em Journal of Econometrics}, 130(1):337--364, 2006.

\bibitem{Hoffman.Gelman.2014}
M.~D. Hoffman and A.~Gelman.
\newblock The no-u-turn sampler: adaptively setting path lengths in hamiltonian
  monte carlo.
\newblock {\em Journal of Machine Learning Research}, 15:1593--1623, 2014.

\bibitem{Joao.2010}
F.~Caldeira. Joao, Márcio~P. Laurin., and S.~Portugal. Marcelo.

\bibitem{MeinholdSingpurwala1983}
Richard~J. Meinhold and Nozer~D. Singpurwalla.
\newblock Understanding the kalman filter.
\newblock {\em The American Statistician}, 37(2):123--127, 1983.

\bibitem{Laurini.2010}
Poletti.~Laurini. Márcio and Koodi.~Hotta. Luiz.
\newblock Bayesian extensions to diebold-li term structure model.
\newblock {\em International Review of Financial Analysis}, 19:342--350, 2010.

\bibitem{Nelson_Siegel_1987}
Charles~R. Nelson and Andrew~F. Siegel.
\newblock Parsimonious modeling of yield curve.
\newblock {\em The Journal of Business}, 60(4):473--489, 1987.

\bibitem{yield_curve_imp}
Barry Nielsen.
\newblock {\em {Bond Yield Curve Holds Predictive Powers} Treasury Rates},
  2017.

\bibitem{Nikolaus.2012}
Hautsch. Nikolaus and Yang Fuyu.
\newblock Bayesian inference in a stochastic volatility nelson–siegel model.
\newblock {\em Computational Statistics \& Data Analysis}, 56:3774--3792, 2012.

\bibitem{Rasmussen2005}
Carl~Edward Rasmussen and Christopher K.~I. Williams.
\newblock {\em Gaussian Processes for Machine Learning (Adaptive Computation
  and Machine Learning)}.
\newblock The MIT Press, 2005.

\bibitem{sambasivan2017statistical}
Rajiv Sambasivan and Sourish Das.
\newblock A statistical machine learning approach to yield curve forecasting.
\newblock {\em IEEE Proc. ICCIDS-2017}, 2017.

\bibitem{sambasivan2018fastGP}
Rajiv Sambasivan and Sourish Das.
\newblock Fast gaussian process regression for big data.
\newblock {\em Big Data Research}, 2018.

\bibitem{Hays_Shen_Huang_2012}
Haipeng~Shen Spencer~Hays and Jianhua~Z. Huang.
\newblock Functional dynamic factor models with applications to yield curve
  forecasting.
\newblock {\em Annals of Applied Statistics}, 6(3):870--894, 2012.

\bibitem{stan_manual}
Stan~Development Team.
\newblock {\em Stan Modeling Language Users Guide and Reference Manual.}, 2016.

\end{thebibliography}

\newpage

\section*{Tables and Figures}

\begin{table}[ht]
\centering
     \begin{tabular}{cl}\hline
                      & Description \\ \hline
                     & \\
        Model 1  &
$(\beta_0, \beta_1, \beta_2,\lambda,\sigma) \sim $ \texttt{Inverge-Gamma(a=1,b=1)}. \\ 
& \\ \hline
&\\
 Model 2        & $(\beta_0, \beta_1, \beta_2) \sim $ \texttt{Gaussian}($0,\sigma_{\beta}$)\\
                    & $(\lambda,\sigma ,\sigma_{\beta})\sim $ \texttt{Inverge-Gamma(a=1,b=1)}. \\\hline

&\\
 Model 3       & $(\beta_0, \beta_1, \beta_2) \sim $ \texttt{Gaussian}($0,\sigma_{\beta}$)\\
                    & $(\lambda,\sigma ,\sigma_{\beta})\sim $ \texttt{Inverge-Gamma(a=0.1,b=0.1)}. \\\hline
     \end{tabular}
\caption{Two different Prior Distribution for Nelson-Siegel Model. Note second and third model allows negative effect over yield on the prior distribution.}\label{tbl_prior_models}
\end{table}

\begin{table}[ht]
\centering
\begin{tabular}{cccccccccccc}\hline
Date       &   1 Mo&	3 Mo& 6 Mo& 1 Yr&	2 Yr&  3 Yr  &	5 Yr  &  7 Yr  &	10 Yr & 20 Yr &	30 Yr\\ \hline
05/01/18&   1.68&	1.85&  2.05&  2.26&	2.50& 2.66  &	2.82 &  2.93  &	2.97	& 3.03 & 3.13 \\
05/02/18&   1.69&	1.84&  2.03&  2.24&	2.49& 2.64 &	2.80	& 2.92  &  2.97	& 3.04 & 3.14 \\
05/03/18&   1.68&	1.84&  2.02&  2.24&  2.49& 2.62 &	2.78	& 2.90  &	2.94	& 3.02 & 3.12 \\
05/04/18&   1.67&	1.84&  2.03&  2.24&	2.51& 2.63 &	2.78	& 2.90 &	2.95	& 3.02 & 3.12 \\
05/07/18&   1.69&	1.86&  2.05&  2.25&  2.49& 2.64&	2.78	& 2.90&	2.95	& 3.02 & 3.12\\
05/08/18&   1.69&	1.87&  2.05&  2.26&	2.51& 2.66&	2.81	& 2.93&	2.97	&3.04 & 3.13\\ \hline
\end{tabular}
\caption{US Treasur yield curve rate ove first six business days of May 2018. \cite{US_treasury_sample_data}}\label{tbl_US_treasury_data}
\end{table}

\begin{figure}[h!]
\centering
\includegraphics[height=4in]{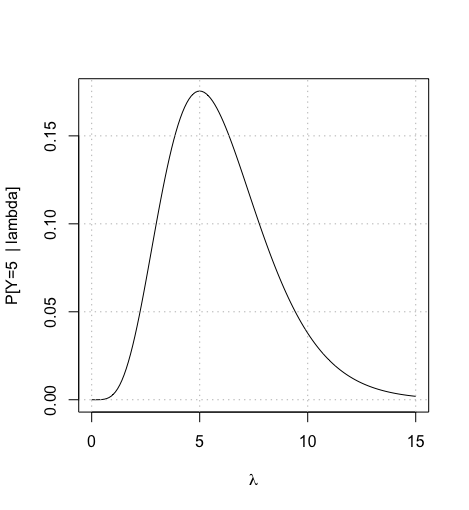}
\caption{The plot shows the most likely value of an unknown parameter $\lambda$, which generates the data $Y=5$. The y-axis represents the likeliness of seeing the data $Y=5$ for different possible values of $\lambda$. The curve is known as the likelihood curve.}
\end{figure}

\begin{table}[ht]
\centering
\begin{tabular}{ccccccc}\hline
                   & $\beta_0$ & $\beta_1$ & $\beta_2$ & $\lambda$ & $\sigma$ & $\sigma_{\beta}$ \\ \hline
Model 1       & 1.639       &    0.255    &   4.831      &    9.052      &   0.143    &     - \\
Model 2       & 3.111       &   -1.440   & -0.016       &   0.950      &   0.043     & 1.636 \\
Model 3       & 3.111       &   -1.440   & -0.012      &   0.954      &      0.036 &   1.705 \\ \hline
\end{tabular}
\caption{The MAP estimates for the Nelson-Siegel parameters, with the three prior distributions in table \ref{tbl_prior_models}. Note that the long-run effect $\beta_0$ for the prior model 1, is less than the medium-term effect $\beta_2$, makes the prior model 1 an undesirable. The similar MAP estimates of the Nelson-Siegel parameters for model 2 and 3 indicates robust posterior analysis, in spite of differences in the prior parameters. }\label{tbl_map_estimates}
\end{table}

\begin{table}
\centering
\begin{tabular}{ccccccc}\\\hline
Parameters & Model  & Mean  &  sd     & $2.5\%$ & Median & $97.5\%$ \\ \hline
$\beta_0$   & m2      &   3.11   &  0.01  & 3.08      &  3.11   &  3.14 \\
                 & m3      &   3.11  &   0.01  & 3.09   &   3.11     &   3.13 \\ \hline
$\beta_1$   & m2      & -1.44  &  0.02  & -1.47  &  -1.44     & -1.40 \\ 
                 & m3      & -1.44   &  0.01  & -1.47  & -1.44     & -1.41  \\ \hline
$\beta_2$  & m2      &   0.00  &  0.16  &-0.33    &  0.00      &   0.32  \\
                 & m3     &   0.01  &  0.14  &-0.27    &  0.01      &   0.27 \\ \hline
$\lambda$ &m2      &   0.97  &  0.12  &   0.76   &   0.96     &   1.23 \\ 
                 &m3      &    0.97 &  0.10  &   0.79  &  0.97    &   1.18  \\ \hline
$\sigma$   &m2      &   0.05  &  0.00  &   0.04  &   0.05   &  0.06  \\
                 &m3&   0.04  &  0.00   &  0.03  &  0.04   &  0.05   \\ \hline
$\sigma_{\beta}$  &m2&   2.32   & 1.18  &  1.11   & 2.02     &   5.43  \\
                &m3&   2.70   & 1.81  & 1.13    & 2.21    &    7.36  \\ \hline
lp        &m2& 155.84 &  1.82 & 151.57 & 156.18 & 158.32 \\
           &m3& 177.40  & 1.80 & 173.18 &177.72 & 179.87 \\ \hline
\end{tabular}
\caption{Monte Carlo estimates of the posterior mean, standard deviation, median, 2.5\% and 97\% quantile of the Nelson-Siegel parameters under model 2 and 3.}\label{tbl_mc_estimate}
\end{table}

\begin{table}
\centering
\begin{tabular}{c|c|c}\\ \hline
Posterior Mean & Posterior Median & 95\% Posterior Confidence Interval \\ \hline
 806.01 &   806.01 & ( 803.59 , 808.46 ) \\ \hline
\end{tabular}
\caption{Poaterior Summary of Bond Price}\label{tabl_posterior_summary}
\end{table}

\end{document}